\documentclass[notitlepage,a4paper,aps,prd,twocolumn,superscriptaddress,nofootinbib,groupedaddress]{revtex4}
\usepackage{graphicx}
\usepackage[colorlinks=true, pdfstartview=FitV, linkcolor=blue, citecolor=red, urlcolor=magenta]{hyperref}

\usepackage{graphics}
\usepackage{amssymb}
\usepackage{amsmath}
\usepackage{url}
\usepackage{color}
\usepackage[utf8]{inputenc}

\usepackage{ulem}
\usepackage{tikz}
\usetikzlibrary{arrows,decorations.markings}

\begin{document}
\newcommand{\cbpf}{
\affiliation{Department of Cosmology, Astrophysics and Fundamental Interactions-COSMO, Centro Brasileiro
de Pesquisas F\'{\i}sicas-CBPF, rua Dr. Xavier Sigaud 150, 22290-180, Rio de Janeiro, Brazil
}
}
\newcommand{\cecs}{
\affiliation{Centro de Estudios Cient\'{\i}ficos (CECs), Av. Arturo Prat 514, Valdivia, Chile, and\\ Universidad San Sebasti\'an, General Lagos 1163, Valdivia, Chile.}
}

\newcommand{\ufes}{\affiliation{PPGCosmo, CCE - Universidade Federal do Esp\'{\i}rito Santo, Av. Fernando Ferrari 514, 29075-910, Vit\'{o}ria-ES, Brazil.}}

\title{Dynamical dimensional reduction in multi-valued Hamiltonians}

\author{Alexsandre L. Ferreira Junior}\email{alexsandre.ferreira@edu.ufes.br}
\ufes
\author{Nelson Pinto-Neto}\email{nelsonpn@cbpf.br}
\cbpf
\author{Jorge Zanelli}\email{z@cecs.cl}
\cecs

\date{\today}

\begin{abstract}
Several interesting physical systems, such as the Lovelock extension of General Relativity in higher dimensions, classical time crystals, k-essence fields, Horndeski theories, compressible fluids, and nonlinear electrodynamics, have apparent ill defined sympletic structures, due to the fact that their Hamiltonians are multi-valued functions of the momenta. In this paper, the dynamical evolution generated by such Hamiltonians is described as a degenerate dynamical system, whose sympletic form does not have a constant rank, allowing novel features and interpretations not present in previous investigations. In particular, it is shown how the multi-valuedness is associated with a dynamical mechanism of dimensional reduction, as some degrees of freedom turn into gauge symmetries when the system degenerates.
\end{abstract}

\pacs{}
\maketitle

\section{Introduction}  

The Hamiltonian formalism unveils important structures of classical mechanical systems through the symplectic geometry of phase space, besides being central in the construction of quantum and statistical mechanics.

Notwithstanding, some systems have apparently ill-defined symplectic structures. Such as the case for Hamiltonians multi-valued in the canonical momenta $p_i$, whose orbits cannot be defined by an initial condition in phase space. Given any point $(q_i,p_i)$, there is more than one value of the Hamiltonian associated with it, consequently, more than one possible time evolution. This phenomenon was spotted in the context of the so-called Lovelock  gravitation theories, the natural extension to higher dimensions of four-dimensional General Relativity \cite{TZ1,TZ2}. The occurrence of this indeterminacy is signaled by the non-invertibility of the Legendre map between the Lagrangian and Hamiltonian representations, $(q, \dot{q}) \rightarrow (q,p)$, and can generically happen in a large class of systems described by an action principle \cite{HTZ1987}. For instance, Horndeski theories, the most general scalar-tensor theories with second order differential equations, possess arbitrary non-trivial kinetic terms, which may present this problem, possibly leading to loss of hiperbolicity in the equations of motion \cite{Bernard2019}, formation of sonic horizons \cite{Gannouji2020}, and appearance of caustics in wave propagation \cite{FKS02,Babichev16}. In the same way, $k-$essence scalar field actions used to source primordial inflation or model dark energy suffer from the same indeterminacy \cite{MN21,JMW07}.

Usually, and in the cases of interest here, the multi-valued character of the Hamiltonian originates from globally non invertible relation between momenta and velocities. The single-valued branches of the function $p_i(q^j, \dot{q}^j)$ are separated by extreme surfaces, where the Hessian determinant $|\partial p_i/\partial \dot{q}^j|$ vanishes.

The equations of motion
\begin{equation}
    \ddot{q}^j \frac{\partial p_i}{\partial \dot{q}^j} +\dot{q}^j \frac{\partial p_i}{\partial q^j} - \frac{\partial L}{\partial q^i}=0,
\end{equation}
develop a singularity at the points where the Hessian degenerates and some of the equations of motion change order. Such singularity is called a degeneracy \cite{saavedra2001} and the locus of points where this occurs defines a surface in phase space, the degeneracy surface.
In this paper we investigate the phase space structure of some multi-valued Hamiltonians as degenerate dynamical systems, which can result in new features and admit different interpretations. In fact, in \cite{Shapere:2012nq, MN21} this singularity in the equations of motion is understood as the existence of some sort of brick wall which inverts the sign of the velocity; whilst in \cite{JMW07} it is a point in which the solution is not determined, referred to as a \textit{terminating singularity}. Here, we see that the presence of a degeneracy can also be interpreted as an irreversible evolution towards --or away from-- the degenerate surface with unbounded velocity or acceleration.

In \cite{wilczek2012, Shapere:2012nq}, Shapere and Wilczek interpret the degeneracy surface as a ground state with a finite velocity, breaking the time symmetry usually present in a standard ground state. In an alternative interpretation, as the system reaches the degeneracy surface, some constraints change from second to first class, which means that some degrees of freedom of the system are no longer dynamical and turn into gauge symmetries \cite{saavedra2001}.

Due to the multi-valuedness of the Hamiltonian, the usual construction of the phase space in terms of coordinates and momenta $(q^i,p_i)$ is obscure. It is therefore useful to work instead with a first order Lagrangian in whose phase space, spanned by the coordinates and velocities as independent variables, the Hamiltonian is a single-valued function. This formalism is reviewed in the next Section, together with a brief summary of dynamical degenerate systems.

In Sec. \ref{sec3}, the discussion in terms of multi-valued Hamiltonians starts with the simplest Lagragian with a non canonical kinetic term. The evolution of the system shows that as it reaches the degeneracy surfaces (when it can), gauge symmetries emerge. The gauge structure at the degenerate surface is analyzed in Sec. \ref{sec4}.

 In Sec. \ref{sec5} we couple the system of Sec. \ref{sec4} to a non degenerate point particle in order to grasp the meaning of having an observable being converted to a gauge mode. In this way, the effects of the degeneracy can be probed by this simple system. We end up with our final remarks in Sec. \ref{sec6}.

\section{Degenerate dynamical systems}  
\label{sec2}

In order to make the connection between a multi-valued Hamiltonian and its degenerate Lagrangian, we shall first present the latter, following closely Ref. \cite{saavedra2001}. Working in a $(2n+1)$ spacetime, we describe the system by a first order Lagrangian one-form $L dt$, where
\begin{equation} \label{lag1}
    L=A_i\dot{z}^i+A_0,
\end{equation}
and $A_\mu\equiv A_\mu(t,z^i)$, which can always be done for any system with a finite number of degrees of freedom \cite{zanelli2008}. The equations of motion are then given through the pre-symplectic form $F_{ij}=\partial_iA_j-\partial_jA_i$ as
\begin{equation}
        F_{ij}\dot{z}^j+E_i=0,
        \label{eqmotion}
\end{equation}
where $E_i=\partial_iA_0-\partial_0A_i$. The velocity is just
\begin{equation}
    \dot{z}^j=-F^{ji}E_i,
    \label{solv}
\end{equation}
$F^{ij}$ the inverse pre-symplectic form where it exists, i.e., $F^{ik}F_{kj}=\delta^i_j$. 

The degeneracy occurs when the two-form $F$ does not have a constant maximal rank $\rho(F_{ij})=2n$ throughout phase space. At some surface $\Sigma$ the Pfaffian $F=\sqrt{\mathrm{det}(F_{ij})}$ vanishes, the rank is reduced to $\rho=2n-2k$, leaving $2k$ indeterminate velocities. Another expression of this problem is that on the degeneracy surface $\Sigma$ the matrix $F_{ij}$ is not invertible and $2k$ of the $2n$ expressions in Eq. (\ref{solv}) become singular, as some components of the inverse $F^{ij}$ diverge.

In the simplest case, $E_i\neq 0$, the phase flow becomes unbounded near the degeneracy surface, and if $F$ has a simple zero the velocity has opposite directions on each side of $\Sigma$. Nonetheless, if $E_i$ also vanishes at a point on $\Sigma$, the equation is identically satisfied, and the flow lines can cross the surface at these points --such case will be discussed later.

Moreover, $\Sigma$ can be characterized by the flux of the Liouville current $j^i= F \dot{z}^i$ through it. Being $n_i=\partial_i F$ the normal to the surface, the flux is
\begin{equation}
    \Phi=j^in_i=-F F^{ij}E_j\partial_i F.
    \label{flux}
\end{equation}
The behavior near the degeneracy surface is defined by the sign of the flux $\eta=\mathrm{sgn}(\Phi)$. For $\eta > 0$, the orbits are outgoing from $\Sigma$ (repulsive degeneracy surface, $\Sigma^{(+)}$), whilst for $\eta < 0$ they are directed towards $\Sigma$ (attractive degeneracy surface, $\Sigma^{(-)}$), with unbounded velocity in both cases. The points on $\Sigma$ where $\Phi$ vanishes correspond to boundaries between attractive and repulsive regions of the degeneracy surface.

Summarizing, the system cannot reach $\Sigma^{(+)}$, it could start there, but then any perturbation would push it away. On the contrary, for configurations near $\Sigma^{(-)}$ are driven towards the degeneracy surface with infinite velocity. Finally, as the surface is reached, the rank of the symplectic form is reduced, turning the system into a constrained one. To better understand the behaviour as the degeneracy is reached, we shall investigate the constraints which arise from the definition of the conjugate momenta $p_i$:
\begin{equation}
    G_i\equiv p_i-A_i\approx0.
\end{equation}
The Poisson brackets between them is just the symplectic form, $\{G_i,G_j\}=F_{ij}$. As the system reaches the degeneracy surface and the symplectic form becomes non invertible, some constraints go from second to first class. Unlike second class constraints, the first class ones give rise to gauge symmetries. Since $F_{ij}$ is antisymmetric its rank is even and if $F$ has a simple zero, two of the $G_i$s among become first class. One of them corresponds to the requirement that the system remains on the degeneracy surface, and one gauge symmetry, generated by the vector tangent to the degeneracy surface appears \cite{saavedra2001}. In the following sections, all the features presented here will appear in a concrete setting of multi-valued Hamiltonians. 

\section{Multi-valued Hamiltonians} 
\label{sec3}
Physical systems in which the momenta are multi-valued functions of the velocities generate degenerate Hamiltonians, as when the kinetic term is a polynomial of degree higher than two in velocities. In order to investigate such situations we start with the simplest Lagrangian with non-canonical kinetic term \cite{HTZ1987,Shapere:2012nq}
\begin{equation}
    L=\frac{\beta}{4}\dot{\phi}^4-\frac{\kappa}{2}\dot{\phi}^2-V(\phi),
    \label{multlagrangian1}
\end{equation}
where $\beta$ and $\kappa$ are positive, and the potential is included so that the orbits fall into the degeneracy surfaces. The conjugate momentum is
\begin{equation}
    p_\phi=\beta\dot{\phi}^3-\kappa\dot{\phi},
\end{equation}
which is a multi-valued function of $\dot{\phi}$. The motion of the system is governed by
\begin{equation}
    (3\beta\dot{\phi}^2-\kappa)\ddot{\phi}=-V'(\phi).
    \label{eqmotion2}
\end{equation}
Here $\dot{\phi}$ and $V'$ denote derivatives with respect to time and $\phi$, respectively. The system becomes degenerate for $\dot{\phi}^2=\kappa/3\beta$, as the coefficient multiplying the acceleration vanishes. Note that the degeneracy surfaces correspond to extremal points of the function $p_\phi(\dot{\phi})$, which separate different single-valued branches.

We can pass from the Hamiltonian to the first order Lagrangian $L=p_\phi(\phi,\rho) \dot{\phi}-H(\phi,\rho)$ in the configuration space spanned by $\vec{z}=(\phi,\rho)$, where $\rho=\dot{\phi}$. The system then is equivalently described as in (\ref{lag1}) by
\begin{equation} \label{L}
L=(\beta \rho^3 - \kappa \rho) \dot{\phi} - \frac{3\beta}{4}\rho^4 +\kappa \rho^2 - V(\phi)\,,
\end{equation}
where we have identified
\begin{align}
A_0&=-H=-\frac{3}{4}\beta\rho^4+\frac{1}{2}\kappa\rho^2-V(\phi)\\
A_1 &=\beta\rho^3-\kappa\rho\;, \qquad  A_2=0\,.
\end{align}
The sympletic form is $F_{ij}=-\epsilon_{ij}\Delta$ with Pfaffian $F=\Delta$, where $\epsilon_{ij}$ is the two-dimensional Levi-Civita symbol, and
\begin{equation} \label{Delta}
    \Delta = 3\beta(\rho-\rho_+)(\rho-\rho_-),
\end{equation}
with $\rho_{\pm}:=\pm\sqrt{\kappa/3\beta}$. Hence, as expected, the degeneracy surfaces are at $\rho_{\pm}$, the zeros of $\Delta$.

Equations (\ref{eqmotion}) take the form
\begin{equation} \label{DegDynamics}
    \Delta\dot{\rho}=-V'(\phi),\hspace{5mm}\Delta(\dot{\phi}-\rho)=0.
\end{equation}
The second identifies $\rho=\dot{\phi}$ when $\Delta\neq0$, and hence the first is just the equation of motion of the system (\ref{eqmotion2}).

\begin{figure}[!ht]
\includegraphics[width=0.5\textwidth]{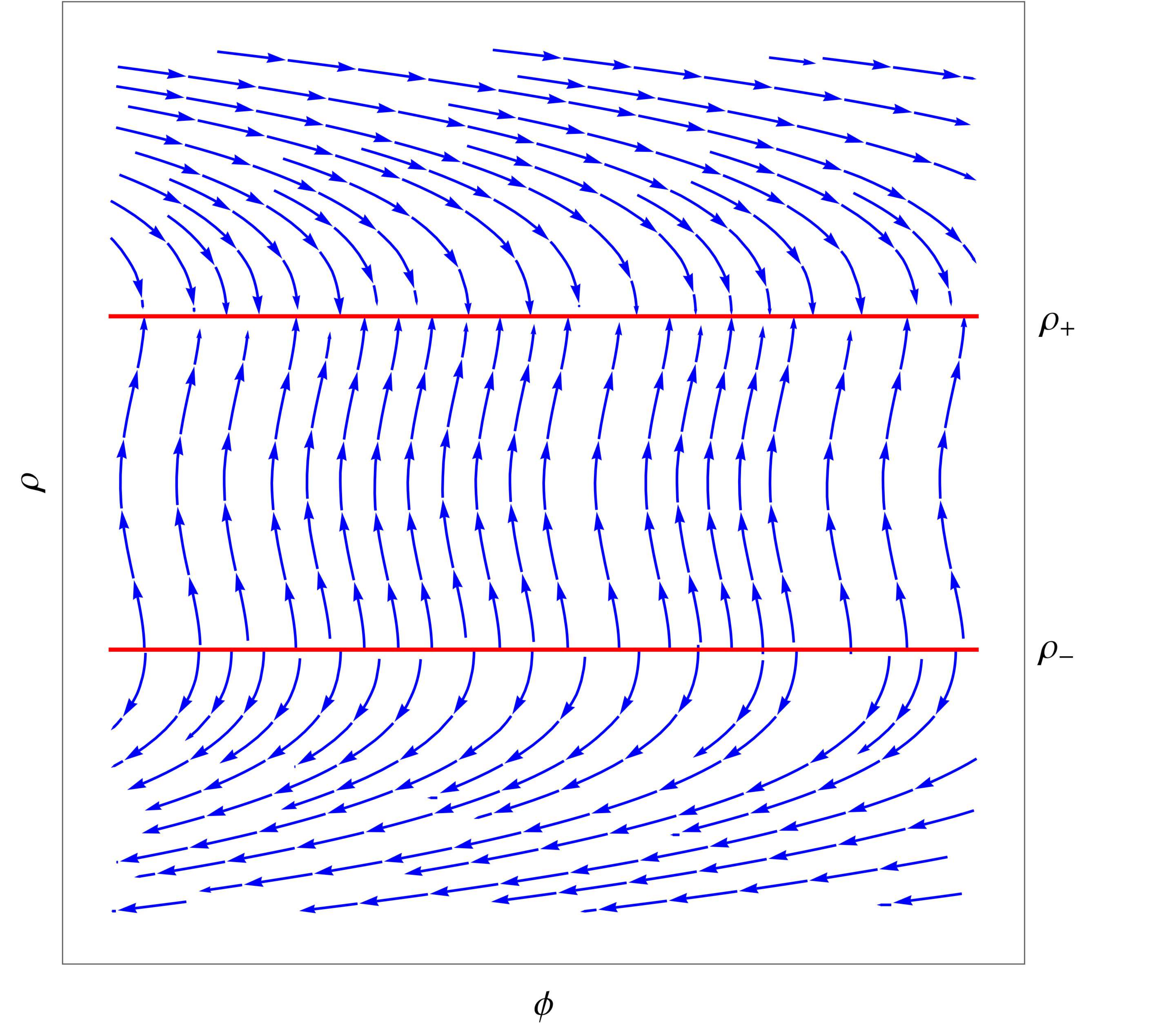}
\caption{Stream plot of the phase space for a linear constant growing potential, depicting the flow of the orbits into $\rho_{+}$ and out of $\rho=\rho_{-}$, respectively.}
\label{fig1}
\end{figure}

From (\ref{flux}), the flux function on the degeneracy surfaces is found to be
    \begin{equation}
        \Phi=-(\partial_i\Delta)\epsilon^{ij}E_j=-6\beta\rho_d V'(\phi_d),
    \end{equation}
defining $\phi_d$ the value of $\phi$ for which the orbit intersects $\Sigma$. Assuming $V'(\phi_d)>0$ --for instance in the case of a linearly growing potential--, $\Phi>0$ for $\rho\rightarrow \rho_{-}$, and $\Phi<0$ for $\rho\rightarrow \rho_{+}$. The stream of the phase flow is presented in FIG.\ref{fig1}, where sign of $\Phi$ indicates that the degeneracy surfaces at $\rho_+$ and $\rho_-$ are attractive and repulsive, respectively.
It is clear from \eqref{DegDynamics} that in the evolution towards $\rho_+$, $\dot{\rho} \rightarrow \pm \infty$, depending on whether the orbit approaches $\rho_+$ from above or below. From the configuration $\rho=\rho_+$ the system can no longer continue in the $(\phi,\rho)$-plane and the evolution stops there. The second equation in \eqref{DegDynamics}, fixes $\dot{\phi}=\rho$ everywhere except at $\rho_{\pm}$, where $\dot{\phi}$ becomes indeterminate.

In \cite{Shapere:2012nq} the degeneracy is interpreted as the presence of a brick wall, that is, the infinite acceleration would reverse the sign of the velocity, as if hitting a wall. This picture would result if the region between the degeneracy surfaces is removed, identifying $\rho_{+}$ and $\rho_{-}$. The system would then reach the degeneracy surface at the point $(\rho_+,\phi_d)$ and instantaneously leave at $(\rho_-,\phi_d)$ with the opposite velocity ($\rho_-=-\rho_+$). 

In \cite{RZ2019} it is shown that degenerate systems such as that in FIG.\ref{fig1} are homotopic to a regular one. That is, a continuous deformation can make the degeneracy disappear. In the example above, as the parameter $\kappa$ approaches zero, the degeneracy surfaces collapse, $\Delta$ acquires a double zero at $\rho=0$ so the system can pass through it and the degeneracy goes away. The case that interests us here is a system in which the degeneracy surfaces are present ($\kappa\neq 0$ above) and $\rho_+ \neq \rho_-$. In this case the system falls into the degeneracy surface and remains trapped there. As will be shown in the next section, once in the surface, the system loses the degree of freedom that corresponds to the motion in the $(\phi,\rho)$-plane which turns into gauge symmetry, so that different points on the degeneracy surface can be the same physical state related by a gauge transformation.
The behavior of the system can be better understood by analyzing its energy function
\begin{equation}
E=\frac{3}{4}\beta\rho^4-\frac{\kappa}{2}\rho^2+V(\phi),
\label{energy}
\end{equation}
which is conserved and therefore each flow line correspond to a constant value of $E(\phi,\rho)$.  Solving \eqref{energy} for $\rho$,
\begin{equation}
\rho^2=\rho^2_d\pm\sqrt{\rho^4_d-\frac{4}{3\beta}(V-E)},
\end{equation}
where $\rho_d$ stands for either one of the roots $\rho_{\pm}$. This last equation can also be rewritten as
\begin{equation}
\rho^2=\rho^2_d\pm\sqrt{\frac{4}{3\beta}\left[V(\phi_d)-V(\phi)\right]}.\label{eq9}
\end{equation}

Considering a more general potential $V(\phi)$, an interesting orbit is one that reaches the degenerate surface at the point $(\phi_d, \rho_d)$ where $\phi_d$ is a critical point of the potential, $V'(\phi_d)=0$. As real roots in \eqref{eq9} require $V(\phi_d)\geq V(\phi)$ for $\phi$ in the vicinity of $\phi_d$, there are no orbits whose value of $\phi_d$ is a minimum of the potential.
   
Moreover, the extreme points of the potential where the flux becomes zero are turning points in the character of the degeneracy surface, from attractive to repulsive, and vice-versa.

\begin{figure}[!ht]
\includegraphics[width=0.5\textwidth]{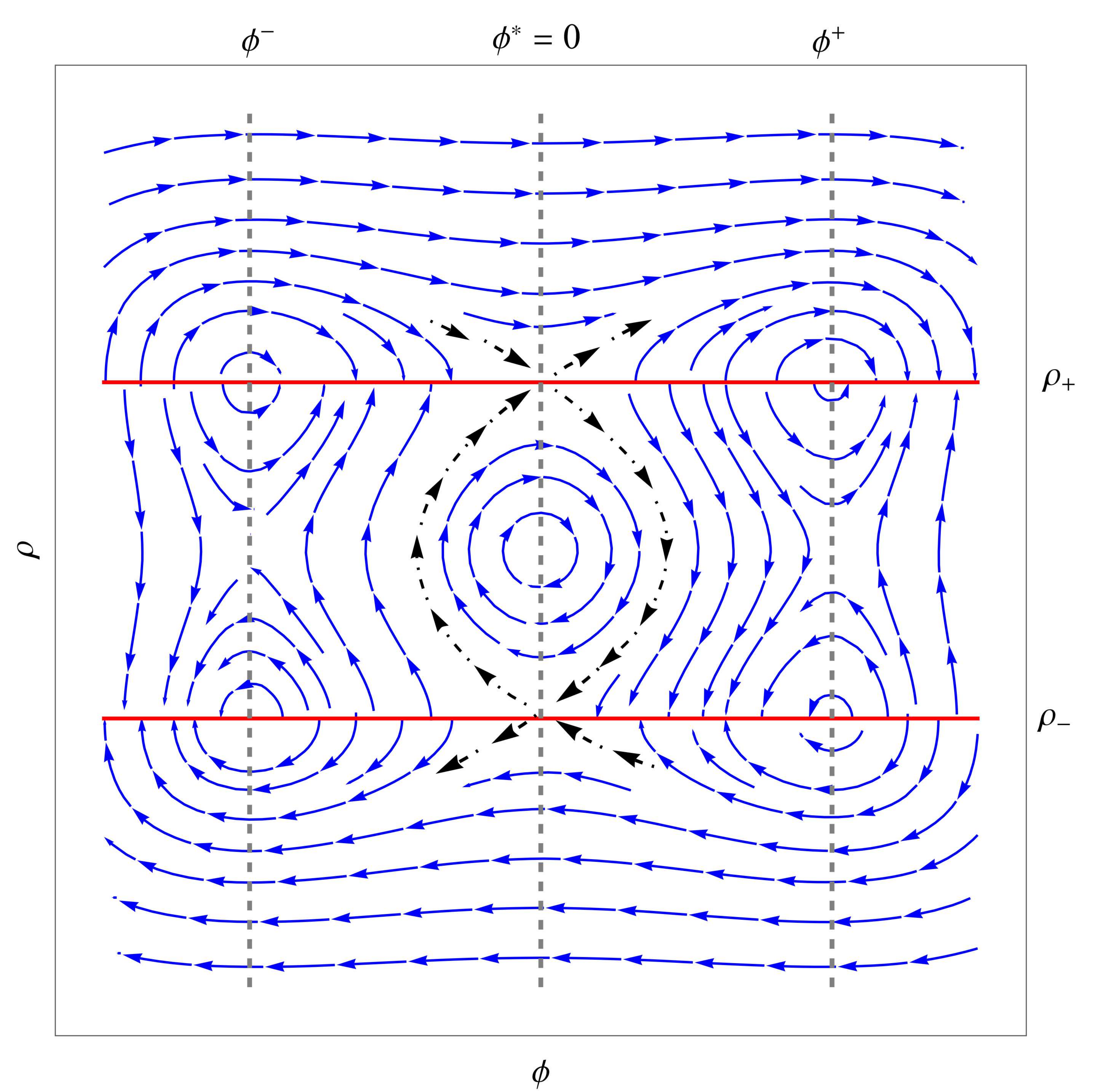}
\caption{Stream plot of the phase space for the quartic potential (\ref{potential}). The dashed gray lines represent the extrema points of the potential, where the degeneracy surfaces change from attractive to repulsive or vice-versa. The dot-dashed black stream lines represent the orbits that can cross the degenerate surfaces at a local maximum of the potential. These lines are separatrixes in the phase space.}
\label{fig2}
\end{figure}

To make the above statements more concrete, consider the potential
\begin{equation}
V(\phi)=\frac{\lambda}{4}\phi^4-\frac{\omega}{2}\phi^2,
\label{potential}
\end{equation}
with $\lambda$ and $\omega$ positive. The potential has two global minima $\phi_{\pm}=\pm\omega/\lambda$, and one local maximum $\phi^{*}=0$ and the flux is
\begin{equation}
\Phi=-6\beta\rho\phi(\lambda\phi^2-\omega).
\end{equation}
At the critical points $\phi_{\pm}, \phi^*$ the flux is zero and the orbits are neither incoming nor outgoing there. However, for the minima $\phi_{\pm}$ there is no orbit approaching the surface, as discussed earlier. The flow in the space $(\phi,\rho)$ is depicted in FIG. \ref{fig2}. The line $\rho=\rho_+$ (the opposite for $\rho=\rho_-$) repeals the orbits with $\phi_d<\phi_{-}$, after which it changes to attractive. Becoming again repulsive for $0<\phi_d<\phi_{+}$, and attractive when $\phi_d>\phi_+$.

The orbits that reach the degeneracy surface at a maximum of the potential are very special in that they are the only ones for which the velocity $(\dot{\phi}, \dot{\rho})$ is always bounded. This can be seen by expanding (\ref{eq9}) around the maximum $V'(\phi_d)=0$ $V(\phi)<V(\phi_d)$, so that the orbit $\rho=\dot{\phi}$ in the vicinity of $\phi_d$ is
\begin{equation}
\dot{\phi}^2=\rho^2_d\pm\sqrt{\frac{2\omega}{3\beta}}(\phi_d-\phi) + O(\phi_d -\phi)^2.
 \label{eq18}
\end{equation}

The crucial thing is that these are the only orbits for which $\dot{\rho}=\ddot{\phi}$ remains finite. This can be seen from \eqref{DegDynamics} and \eqref{eq9} using $\Delta= 3\beta(\rho^2-\rho_d^2)$ and expanding $V'(\phi)$ around $\phi_d$, which yields
\begin{equation}
\dot{\rho}= \sqrt{-V''(\phi_d)/(24\beta)}(\phi_d -\phi) + O(\phi_d -\phi)^2 \,.
\end{equation}
(This is well defined for $V$ having a maximum at $\phi_d$ and therefore $V''(\phi_d)<0$).

For the case when $\dot{\phi}^2 \geq \rho^2_d$ we have that $\phi$ can be positive or negative. First, when $\phi>0$, eq. (\ref{eq18}) can be solved to give
\begin{equation}
\phi=\sqrt{\frac{\omega}{24\beta}}t^2-\sqrt{\frac{3\beta}{2\omega}}\rho^2_d.
\end{equation}
The orbit crosses the degeneracy surface for two values of $t=t_\pm=\pm\sqrt{6\beta/\omega}\rho_d$. As $t\rightarrow t_-$, $\phi$ goes from positive to zero and $\dot{\phi}\rightarrow\rho_-$. Then, as $t$ grows, $\dot{\phi}$ becomes positive and $\phi$ starts to grow, reaching again $\phi=0$ for $t=t_+$ with $\dot{\phi}=\rho_+$.

Summing up, for $\dot{\phi}^2\geq\rho^2_d$ and $\phi>0$, the orbit approaches $(0, \rho_-)$, crossing the degeneracy surface from below. Then reaching $(0, \rho_+)$ from below and $\phi<0$, crossing the $\rho_+$ line at $\phi=0$, to end up finally trapped at $(\phi_f, \rho_+)$, for some $\phi_f>\phi_+$. An analogous evolution occurs for the orbit approaching the maximum at $(0,\rho_+)$ from the $\phi<0$, $\rho> \rho_+$. This is described by the two dot-dashed black lines in FIG. \ref{fig2}, which intersect each other.  This intersection reflects the unstable nature of the orbits that pass over a local maximum, for which any infinitesimal perturbation changes the topology of the orbit.

These curves are separatrixes in the space $(\phi,\rho)$, dividing the region between $\rho_+$ and $\rho_-$ into one in which the orbits never reach the degeneracy surfaces, and those that end on them.

It is important to remark that in the orbits described above, the system is able to cross the degeneracy surface without losing degrees of freedom, because there is no meaning in defining the degrees of freedom of a single point in the orbit. Thus, as the orbits crosses the surface, all dynamical variables and its derivatives are continuous and finite and do not get trapped, in contrast with previous cases. The system then only lose degrees of freedom later, when it falls into the degeneracy surface in a point into which the flux is negative, remaining there.
%

\section{Emergence of gauge symmetry}  
\label{sec4}
Attractive degenerate surfaces are of special interest, as the system can fall into them, becoming degenerate. As we shall see, at those surfaces of the phase space some degrees of freedom of the system turn into gauge parameters.

A typical symptom of gauge invariance is that the dynamical equations do not determine the time evolution of a certain coordinate, which can therefore take arbitrary values at any time. This coordinate is identified as a gauge parameter rather than a dynamical variable. This is exactly what happens with \eqref{DegDynamics}, which leaves the evolution equation for $\phi$ indeterminate if $\Delta=0$.

The system described by the first order Lagrangian \eqref{L} has the following constraints
\begin{equation}
G_1=p_\phi-\beta\rho^3+\kappa\rho\approx0,\hspace{5mm}G_2=p_\rho\approx0.
\end{equation}
Their Poisson bracket,
\begin{equation}
\{G_1,G_2\}=-\Delta,
\end{equation}
is generically nonzero, but it vanishes at the degeneracy surface $\Delta=0$. Following the Dirac-Bergmann procedure, we write the total Hamiltonian as
\begin{align}
H_T&=p_\phi\dot{\phi}+p_\rho\dot{\rho}-L+\mu^iG_i\\
&=(p_\phi-A_1)\dot{\phi}+p_\rho\dot{\rho}+H+\mu^iG_i \\
&=\frac{3\beta}{4}\rho^4-\kappa \rho^2 +V(\phi) + \bar{\mu}^iG_i.
\end{align}
The consistency conditions $\dot{G}_i= \{G_i,H_T\} \approx0$ give
\begin{align}\label{consist1}
\dot{G}_1 &=-V'(\phi)-\Bar{\mu}^2\Delta\approx0, \\
\dot{G}_2 &=\Delta(\Bar{\mu}^1-\rho)\approx0,
\label{consist2}
\end{align}
which determine the Lagrange multipliers everywhere, --except on the degeneracy surfaces--, there are no new constraints and the Dirac procedure stops here. There is still the pending issue of what can be said about the system at the degeneracy surface. 

The equations of motion are
\begin{equation}
\dot{\phi}=\{\phi,H_T\}=\Bar{\mu}^1,\hspace{5mm}\dot{\rho}=\{\rho,H_T\}=\Bar{\mu}^2.
\end{equation}
The evolution of the system, therefore, is given by the Lagrange multipliers. As the orbits approach the degeneracy surface, one would naively expect that $\dot{\rho}=\Bar{\mu}^2\approx -V'(\phi)/\Delta$ diverges in order for (\ref{consist1}) to be satisfied. On the other hand, $\Bar{\mu}^1$ becomes an arbitrary function of time, which is consistent with the interpretation of $\phi$ becoming a gauge parameter at the degeneracy surface.

The constraints $G_1$ and $G_2$ generate translations in $\phi$ and $\rho$, respectively. At the degeneracy surface, these generators become first class and can be interpreted as generators of some gauge transformations. However, the demand of being at the degeneracy surface yields a new constraint, namely,
\begin{equation} \label{gauge.fixing}
\varphi\equiv \rho-\rho_d \approx 0\,.
\end{equation}
This expression can be recognized as a gauge fixing condition for the gauge generator $G_2$, while it has vanishing Poisson bracket with $G_1$. Hence, the system, once trapped at the degeneracy surface, has three constraints, $G_1\approx 0$, $G_2\approx 0$ and $\varphi\approx 0$. The first is a genuine first class generator, while the other two form a pair of second class constraints that reduce the phase space by two dimensions. At the degeneracy surface, $G_1$ generates translations along $\phi$, which could now be seen as a gauge direction. Therefore, as the system falls into the degeneracy surface it stays trapped there and has no remaining degrees of freedom. This is different from the interpretation of \cite{Shapere:2012nq}, where the degeneracy surface is viewed as a ground state with a finite velocity, which therefore breaks the continuous time-translation symmetry.

--------\\
There is yet one important remark; In in the case with potential (\ref{potential}), the attractive or repulsive character of the degeneracy surface changes at the critical points $V'(\phi)=0$. Therefore, one can imagine that after the system falls in the attractive part of the degeneracy surface, a gauge transformation could change $\phi$ into the region where the surface becomes repulsive, from where it could be expelled. This is a problem because, as all points represent the same physical state, one cannot define from which point of the surface the system would emerge.

However, once the system falls into the surface, it changes completely, losing all degrees of freedom. Having no dynamics anymore, it can no longer be described by the previous equations of motion. An alternative way to see this situation is to consider that reaching the degeneracy surface at an attractive point $\rho$ becomes fixed at the value $\rho_d$. The Lagrangian would no longer be given by \eqref{L}, but by
\begin{equation}
L'= \frac{2\kappa}{3}\rho_d\,\dot{\phi} + \frac{\kappa^2}{4\beta}-V(\phi),
\end{equation}
which does not describe a dynamical system: the first term is a total derivative and therefore does not provide an equation of motion for $\phi$. Extremizing the action with respect to $\phi$ merely says that the system will stay at one of the critical points of $V$. Of course this is just the classical analysis; quantum mechanically, the system will surely have more interesting features \cite{deMicheli}.
This alternative description also dispels some possible ambiguity due to the fact that the Legendre transformation that maps the original Lagrangian to the first order formalism, is not globally invertible, which could undermine the analysis made in this section. Nevertheless, once understood that the system gets trapped at the attractive degeneracy surface, the conclusions found become clear, even from the original dynamical equation \eqref{eqmotion2}. Namely, that the only prediction about the system is that it is at the degeneracy surface, the motion tangent to the surface being spurious.

\section{Probing the degeneracy}  
\label{sec5}

In order to analyze the emergence of gauge symmetries at the degeneracy surfaces, we couple a degenerate system with the simplest non-degenerate one: a point particle. The coupling is with the time derivative of $\phi$, because as the field reaches the degeneracy surface, $\dot{\phi}$ has a well defined limit while $\phi$ becomes a gauge parameter. We take the Lagrangian as
\begin{equation}
L=\frac{\beta}{4}\dot{\phi}^4-\frac{\kappa}{2}\dot{\phi}^2-V(\phi)+\frac{m}{2}\dot{x}^2+q\dot{\phi}x.
\end{equation}
The conjugate momenta are
\begin{equation} \label{momenta}
p_\phi=\beta\dot{\phi}^3-\kappa\dot{\phi}+qx,\hspace{5mm}p_x=m\dot{x},
\end{equation}
and the equations of motion
\begin{equation}
(3\beta\dot{\phi}^2-\kappa)\ddot{\phi}=-V'(\phi)+q\dot{x},\hspace{5mm}m\ddot{x}=-q\dot{\phi}.
\end{equation}
The degeneracy surfaces, where the equation of motion for $\phi$ breaks down, remain the same as before, $3\beta\dot{\phi}^2-\kappa=0$. The particle is driven by a force proportional to $\dot{\phi}$ which is not constant in time if $V'\neq0$. However, as we saw, when the system reaches the surface $\Delta=0$ and gets stuck there, $\dot{\phi}$ is frozen, and the force on the particle is now constant. The equation for the particle is integrated to give
\begin{equation}
m\dot{x}+q\phi=\mathrm{const}=am/q\,,
\label{solpar}
\end{equation}
where $a$ is a constant of integration. Substituting in the equation of motion for $\phi$, we have again eq. (\ref{eqmotion2}), however subjected to an effective potential $V_{\textrm{eff}} = V(\phi)+ q^2\,\phi^2/(2m)- a\phi$. The situation for the degenerate subsystem is analogous to the previous case, although the system now has an effective potential which depends on the initial velocity of the probe particle.

The Hamiltonian of the system can be written as
\begin{equation}
H_c=\frac{3\beta}{4}\dot{\phi}^4-\frac{\kappa}{2}\dot{\phi}^2+V(\phi)+\frac{m}{2}\dot{x}^2\,,
\label{coupledhamil}
\end{equation}
where $\dot{\phi}$ is a function of $p_\phi$ implicitly given by the inverse of \eqref{momenta}. Clearly, as the inverse $p_\phi (\dot{\phi})$ is multivalued, so is the Hamiltonian.

\begin{figure}[!ht]
\includegraphics[width=0.5\textwidth]{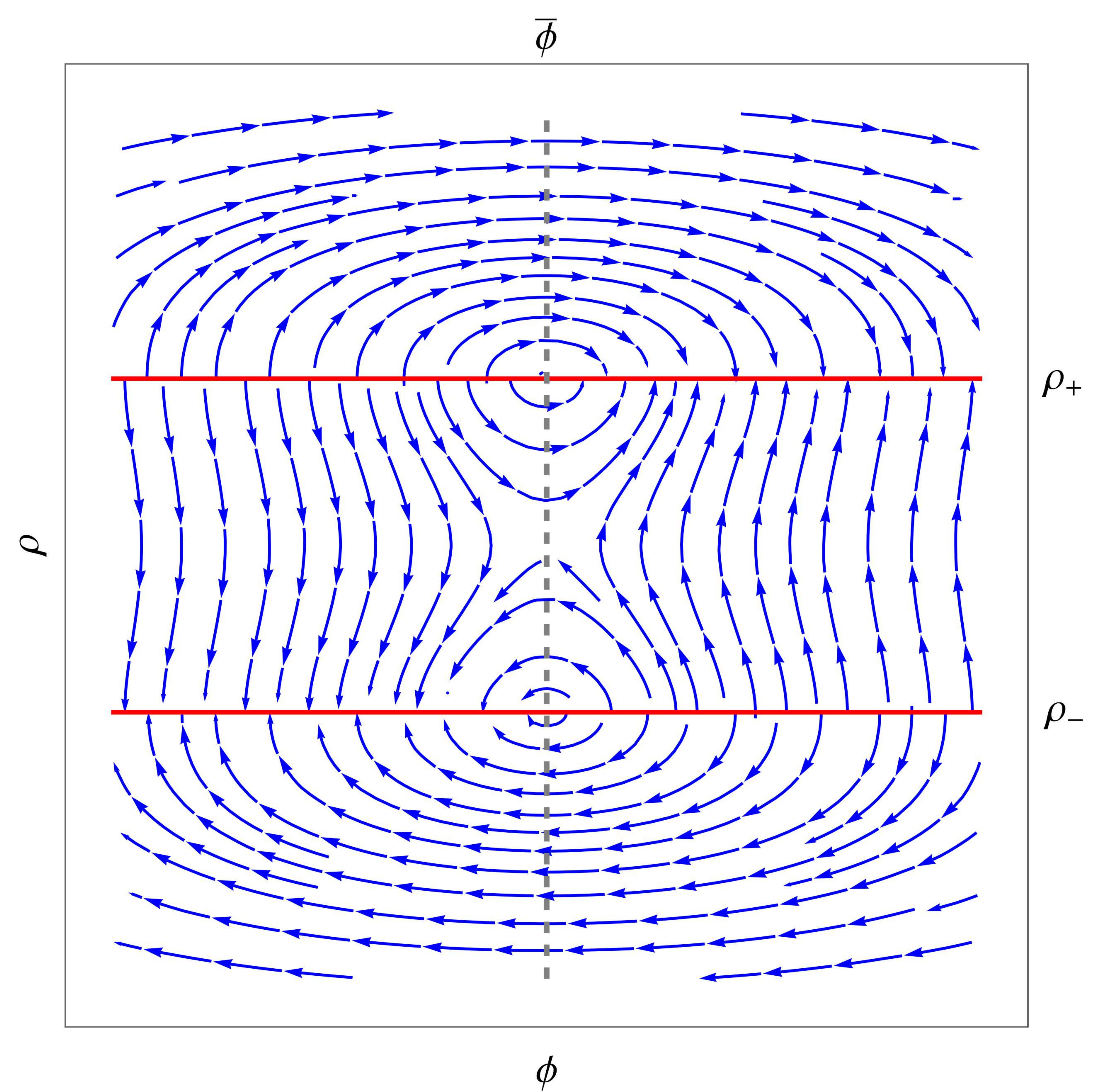}
\caption{Stream plot of the phase space $(\phi,\rho)$ for a linear potential $V(\phi) \propto \phi$, coupled to a regular point particle. The coupling introduces a value of $\phi$ indicated by the dashed gray line, which depends on the initial condition for $\dot{x}$, where the degeneracy surface changes character.}
\label{fig3}
\end{figure}

On the other hand, in the first order formalism, with coordinates $z^i=(\phi,\rho,x,v)$, and
\begin{equation}
A_1=\beta\rho^3-\kappa\rho+qx,\hspace{2mm}A_2=0,\hspace{2mm}A_3=mv,\hspace{2mm}A_4=0,
\end{equation}
and $A_0=-H_c$. The first order system now has constraints
\begin{align}
G_1 &=& p_{\phi}-\beta \rho^2 +\kappa \rho -qx \approx 0, \quad G_2 = p_{\rho}\approx 0,   \\
G_3 &=& p_x -mv \approx 0, \quad \quad \qquad \qquad G_4 = p_v \approx 0.
\end{align}
The symplectic form $F_{ij}=\partial_i A_j - \partial_j A_i$ reads
\begin{equation}
[F_{ij}]=\begin{bmatrix} 
0 & -\Delta & -q & 0 \\
\Delta & 0 & 0 & 0\\
-q & 0 & 0 & -m  \\
0 & 0 & m & 0 
\end{bmatrix}\;,
\end{equation}
where $\Delta=3\beta \rho^2-\kappa$, as before. The determinant $\mathrm{det}(F_{ij})=\Delta^2m^2$, degenerates for $\Delta=0$, and the first order equations of motion read
\begin{align}\label{eq52}
\Delta\,\dot{\rho}&= -V'+q\dot{x}\,,\qquad \qquad \Delta(\dot{\phi}-\rho)=0\,,\;\\
\label{eq53}
m\,\dot{v}&= -q\dot{\phi}\, ,\qquad \qquad \qquad m(\dot{x}-v)=0\;.
\end{align}
It is clear from eq. (\ref{eq52}) that the character of the degeneracy surface depends on the velocity $v_d$ of the particle when it intersects $\rho^2=\rho^2_d$. One can see that also through the flux at the degeneracy surface
\begin{equation}
\Phi=m^2(-V'+q\dot{x})6\beta\rho.
\end{equation}
The flux vanishes for $\dot{x}=V'/q$, i.e. at $\phi=\bar{\phi}=m(a-V')/q^2$, where the character of the degeneracy surface changes between attractive and repulsive. This is depicted in FIG. \ref{fig3} for a potential linear in $\phi$, in which case the points ($\phi=\bar{\phi}, \rho=\rho_d$) correspond to fixed points of the Hamiltonian flow.

The coupled system illustrates well the occurrence of the degeneracy. Suppose the system starts at a generic point in the 4-dimensional phase space. The evolution is determined by the autonomous system
\begin{align}
\dot{\phi} &= \rho \;,\qquad \qquad \dot{\rho}=\frac{qv_0-V'(\phi)- q^2(\phi-\phi_0)/m}{3\beta \rho^2 - \kappa} \\
\dot{x}&= v\;, \quad \quad \qquad \dot{v}=-q \rho/m \;,
\end{align}
where the dependence of $\rho$ on the initial values $\phi_0$ and $v_0$ are explicitly shown. In principle, the equations for $\phi$ and $\rho$ can be integrated (numerically perhaps). One can assume this has been achieved and therefore $\phi(t)$ and $\rho(t)$ are known functions that depend on the initial configuration. Then, equations for the remaining phase space coordinates $(x,v)$ can be directly integrated. So long as the evolution does not reach the degeneracy surfaces $\rho=\rho_{\pm}$, the solution takes the form  
\begin{align}\label{rho(t)}
\rho(t) &=F(t;\phi_0,\rho_0,v_0)\, ,\\ \label{phi(t)}
\phi(t) &= \int_0^t \rho(t') dt' + \phi_0\, ,\\ \label{v(t)}
v(t) &=-\frac{q}{m}(\phi(t)-\phi_0) + v_0\, ,\\ \label{x(t)}
x(t) &= \int_0^t v(t') dt' + x_0\,.
\end{align}

Suppose a generic case, in which the system reaches the degeneracy surface $\rho=\rho_d$ in a finite time, $t=t_d$. Then, for $t>t_d$ the equations for $(x,v)$ become those of a uniformly accelerated, non-degenerate, particle,
\begin{equation} \label{v-x}
\dot{v}=-\frac{q}{m}\rho_d\,, \qquad \dot{x}= v(t)\,, \quad \mbox{for} \quad t>t_d\,,
\end{equation}
whose solution is
\begin{align}
x(t) = x_d \, +\, \left[v_0- \frac{q}{m}(\phi_d-\phi_0)\right]\, t \,-\,\frac{q\rho_d}{2m}\, t^2 \,,
\end{align}
where $\phi_d$, and $x_d$ are given by (\ref{phi(t)},\ref{x(t)}) evaluated at $t=t_d$, while the sub space $(\phi,\rho)$ reduces to the point $(\phi_d, \rho_d)$. Hence, following the point particle trajectory leads to distinct sets of information about the dynamical system when $t<t_d$ and $t>t_d$. When $t<t_d$, ($\phi_0,\rho_0,v_0$) determines completely $\rho(t)$, see Eq.~\eqref{rho(t)}, which in turn determines $\phi(t),v(t)$ and, together with $x_0$, $x(t)$. In this case, inspecting a particular point particle trajectory at $t<t_d$ allows the determination of all these initial constants. However, after the system reaches the degeneracy surface at $t>t_d$, the point particle trajectory follows uniformly accelerated trajectories, from which one can only determine $(x_d,v_d = v_0 - q(\phi_d-\phi_0)/m,\rho_d)$, and complete information about the original physical system and its initial constants is lost. It is impossible to disentangle them if one has no access to the non-degenerate era. Henceforth, just as in the coupled system proposed in \cite{saavedra2001}, after the system reaches the degeneracy surface, the information about the $t<t_d$ era is essentially lost.

\section{Final Remarks}
\label{sec6}
    
 The multi-valuedness of the Hamiltonian with respect to the canonical momenta due to a non invertible Legendre map gives rise to an ambiguous dynamical evolution.

As shown in \cite{saavedra2001}, a consistent interpretation can be given in which the system loses degrees of freedom at the degeneracy surfaces in an irreversible manner. This is also seen in the examples discussed here, which in spite of the ill-defined Legendre map, can always be expressed as first order systems, in which position and velocity are treated as independent dynamical variables. This approach allows a more controlled analysis of the evolution of the dynamical system towards a degenerate surface. In particular, it is clearly seen that when the orbit falls into the surface, the system is trapped and loses the corresponding degree of freedom. If the system combines degenerate and non-degenerate sectors, different orbits starting from different initial states that fall into the same degeneracy surface have the same evolution equations afterwards. The only memory of the degenerate prehistory would be contained in the initial conditions of the subsequent non degenerate evolution. However, the information about the system before the degeneracy is reached is inaccessible due to the impossibility to discern it from the choice of initial conditions on the post-degenerate system. Hence, multi-valued Hamiltonian systems provide a dynamical dimensional reduction mechanism.

This analysis can possibly be used in a plethora of interesting physical situations. In fact, more general multi-valued systems were proposed in Ref. \cite{Shapere:2012nq}: The ``fhg" model, whose kinetic term coefficients are functions of $\phi$; and the ``double sombrero" model, a two-dimensional system in which
the potential and the kinetic term are both rotationally invariant quartic polynomials in the $(\psi_1;\psi_2)$ and $(\dot{\psi}_1; \dot{\psi}_2)$ planes, respectively. In the former, the degenerate surface depends on the choice of coefficients, enabling more interesting shapes and even a compact surface in phase space rather than a line. Whilst the double sombrero is a generalization of the model discussed here, having more degrees of freedom (albeit both pairs of conjugate coordinates degenerate in the same surfaces, remaining no degrees of freedom of the system to probe the
transition).

Such models are well suited to investigate the consequences of the degeneracy in physically interesting situations. For example modelling Horndeski theories, cosmological $k$-essence models, and some kind of spontaneous symmetry breaking in the degeneracy surface. Notwithstanding, the general features of the degeneracy goes in the same lines as discussed here: As it degenerates, the systems loses degrees of freedom, emerging as a different lower dimensional dynamical system, with no memory of its past.

\begin{acknowledgments}
NPN acknowledges the support of CNPq of Brazil under grant PQ-IB 310121/2021-3. ALFJ was supported by the Research Support Foundation of Esp\'{\i}rito Santo - FAPES grant number 13/2019. JZ's research has been partially supported by grants 1180368 and 1220862 from FONDECYT/ANID.
\end{acknowledgments}

\end{document}